\documentclass[reprint,prl,aps,epsfig,longbibliography,superscriptaddress,twocolumn]{revtex4-1}
\usepackage{graphicx}
\usepackage[unicode=true,colorlinks=true]{hyperref}
\hypersetup{linkcolor=blue,citecolor=blue,urlcolor=blue}
\usepackage{dcolumn}
\usepackage{bm}
\makeatletter
\usepackage{color, colortbl}
\definecolor{Gray}{gray}{0.9}

\newcommand{\Rmnum}[1]{\expandafter\@slowromancap\romannumeral #1@}

\begin{document}
\title{Magnetic field dependent equilibration of fractional quantum Hall edge modes}
\begin{abstract}
Fractional conductance is measured by partitioning  $\nu = 1$ edge state using gate-tunable fractional quantum Hall (FQH) liquids of filling 1/3 or 2/3 for current injection and detection. We observe two sets of FQH plateaus 1/9, 2/9, 4/9 and 1/6, 1/3, 2/3 at low and high magnetic field ends of the $\nu = 1$ plateau respectively. The findings are explained by magnetic field dependent equilibration of three FQH edge modes with conductance $e^2/3h$ arising from edge reconstruction. The results reveal remarkable enhancement of the equilibration lengths of the FQH edge modes with increasing field.
\end{abstract}

\author{Tanmay Maiti}
\affiliation{Saha Institute of Nuclear Physics, HBNI, 1/AF Bidhannagar, Kolkata 700 064, India}
\author{Pooja Agarwal}
\affiliation{Saha Institute of Nuclear Physics, HBNI, 1/AF Bidhannagar, Kolkata 700 064, India}
\author{Suvankar Purkait}
\affiliation{Saha Institute of Nuclear Physics, HBNI, 1/AF Bidhannagar, Kolkata 700 064, India}
\author{G J Sreejith}
\affiliation{Indian Institute of Science Education and Research, Pune 411008, India}
\author{Sourin Das}
\affiliation{Department of Physical Sciences, IISER Kolkata, Mohanpur, West Bengal 741246, India}
\author{Giorgio Biasiol}
\affiliation{Istituto Officina dei Materiali CNR, Laboratorio TASC, 34149 Trieste, Italy}
\author{Lucia Sorba}
\affiliation{NEST, Istituto Nanoscienze-CNR and Scuola Normale Superiore, Piazza San Silvestro 12, I-56127 Pisa, Italy}
\author{Biswajit Karmakar}
\email{biswajit.karmakar@saha.ac.in}
\affiliation{Saha Institute of Nuclear Physics, HBNI, 1/AF Bidhannagar, Kolkata 700 064, India}

\pacs{ 75.47.Lx, 75.47.-m} \maketitle
%\keywords{IQHE,FQHE,}
\maketitle

Topological phases of matter such as the quantum Hall states possess gapless protected surface states that carry the current, while the bulk remains insulating. At the smooth boundary Coulomb interaction leads to reconstruction of the edge states in integer quantum Hall (IQH) systems \cite{Beenakker1990,Wen1994,Johnson1995}, fractional quantum Hall (FQH) systems \cite{PhysRevLett.88.056802,PhysRevLett.111.246803,PhysRevB.68.125307,PhysRevB.68.035332,PhysRevLett.91.036802}, quantum spin Hall insulators \cite{Wang2017,Amaricci2017} and graphene nanoribbons \cite{Ihnatsenka2013} etc. 
Edge reconstruction \cite{MacDonald1990,meir1994,Bid2010,Inoue2014} can result
in additional integer and fractional edge modes as well as neutral modes with short equilibration lengths \cite{Chklovskii1992,Wang2017,Amaricci2017,Ihnatsenka2013} due to quasi-particle scattering between the modes. Transient FQH edge modes with short equilibration lengths of a few micrometers have been demonstrated in IQH \cite{Kouwenhoven1990} and FQH \cite{sabo2017} systems.
Robust reconstructed FQH edge modes with long equilibration lengths \cite{Ronen2018,Nakamura2019} can have non-trivial implications on investigating braiding statistics \cite{Halperin1984,Nayak2008,Nakamura2019,Kim2006,Arovas1984}, quantum interferometry \cite{McClure2012,Ofek2010,Park2015} and design of hybrid quantum Hall systems \cite{alicea2016topological}. 
A suitable system to explore robust FQH edge modes is $\nu = 1$ IQH state, where three edge modes with conductance $e^2/3h$ \cite{Beenakker1990} are formed by edge reconstruction. Robustness of the FQH edge modes is investigated here for the first time by varying magnetic field.

In this letter we show enhancement of equilibration length of a FQH edge mode of conductance $e^2/3h$ upon increasing magnetic field within the $\nu = 1$ IQH plateau. The equilibration length of the mode is estimated to be as high as $777 \pm 40 \  {\rm \mu m}$ at the high field end of the plateau. Enhancement of the equilibration length with increasing field is corroborated by investigating equilibration length of FQH edge modes at bulk filling $\nu = 2/3$. Our results reveal the way to find the robust FQH edge modes.

\begin{figure}[h!]
\includegraphics[width=\columnwidth]{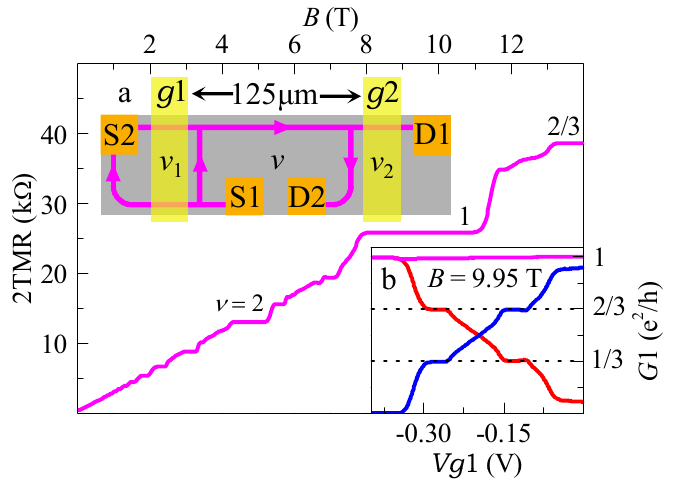}
\centering \caption[ ] 
{\label{fig:device diagram} (color online) Two terminal magneto-resistance (2TMR) trace. Inset-a: Topologically equivalent device structure. Magenta line shows the current path connecting the ohmic contacts. Inset-b: Characteristics of the top gate $g1$ at $9.95 \ {\rm T}$. Blue and red curves represent transmitted (${\rm S1} \rightarrow {\rm S2}$) and reflected (${\rm S1} \rightarrow {\rm D2}$) conductance respectively and magenta line represents sum of those two conductance.}
\end{figure}

The experiments are carried out on a modulation doped heterostructure, in which the 2DEG resides in a GaAs/AlGaAs heterointerface located $100 \ {\rm nm}$ below the top surface. Figure \ref{fig:device diagram} (inset a) shows a topologically equivalent schematic device structure \cite{SM2}, where the two sides of IQH region of filling $\nu$ is bounded by regions of filling fractions $\nu_1$ and $\nu_2$ that are tunable using top-gates $g1$ and $g2$ as in Ref.~\onlinecite{karmakar2011controlled}. For transport measurements, four standard Ohmic contacts (S1,S2,D1,D2) are deployed on the device. The sample is mounted in a dilution refrigerator equipped with $14 \ {\rm T}$ superconducting magnet at base temperature $7 \ {\rm mK}$, where the lowest electron temperature achieved is about $30 \ {\rm mK}$. All measurements are carried out at $30 \ {\rm mK}$ unless stated otherwise. The carriers are injected by light illumination at $3 \ {\rm K}$ with a GaAs LED and the injected carriers are persistent at low temperature \cite{nathan1986persistent}. After illumination, the sample carrier density and mobility become $n \sim 2.27\times10^{11} \ {\rm cm^{-2}}$ and $\mu \sim 4\times10^6 \ {\rm cm^2/Vs}$ respectively. At S2, a customized preamplifier SR555 (the RC filter is  removed from the bias input) is deployed to facilitate simultaneous measurement of the output current and application of AC excitation voltage. Source contacts S1 and S2 are excited by $25.8 \ {\rm \mu V}$ at frequency $17 \ {\rm Hz}$ and $26 \ {\rm Hz}$ respectively \cite{SM2}, such that the excitation injects a system current of $1 \ {\rm nA}$ corresponding to a quantized conductance of $e^2/h$. Output currents at D1 and D2 are measured by lock-in technique using suitable current to voltage preamplifiers. 

To set the filling fraction $\nu = 1$ of the 2DEG, two terminal magneto-resistance (2TMR) is measured between contacts S1 and D2 (Fig. \ref{fig:device diagram}) disconnecting all other contacts and setting gate bias $Vg1 = Vg2 = 0$ V \cite{SM2}. A $\nu = 1$ IQH plateau is formed in the magnetic field $B$ range of $8$ to $11 \ {\rm T}$.
Figure \ref{fig:device diagram} (inset b) shows the transmitted conductance between S1 to S2 (blue curve) and the reflected conductance between S1 to D2 (red curve) at $B = 9.95 \ {\rm T}$ keeping the top gate $g2$ in pinch-off condition. This $g1$ gate transmission characteristic shows incompressible FQH plateaus at filling fractions $\nu_1 = 1/3$ and $2/3$ similar to Ref.~\onlinecite{grivnin2014nonequilibrated}. The transmitted conductance does not reach $1$ at $Vg1 = 0 \ V$, because of insufficient electron density beneath the top gates. Total conductance stays fixed (magenta line) at unity as expected from current conservation. The characteristics of both $g1$ and $g2$ gates are identical and the positions of $2/3$ and $1/3$ plateaus shift to higher gate voltage bias with increasing $B$ field \cite{grivnin2014nonequilibrated} to achieve the same incompressible state.

In our experiments, the current from the source S2 transmitted through the $\nu_1$ FQH fluid and the current from the source S1 reflected by the $\nu_1$ fluid, flow along the top mesa boundary (Fig. \ref{fig:device diagram} inset a). At the detector side, the transmitted current through  $\nu_2$ FQH fluid reaching D1 and the current reflected by $\nu_2$ FQH fluid reaching D2 are measured. Corresponding low frequency (DC) two-terminal combined conductance (TTCC) is denoted by $G^{~\nu_1,\nu_2}_{\rm S \rightarrow D}$ incorporating all the conditions of measurements, where $\rm S$($\rm D$) denotes the source(detector) contact.

\begin{figure}[h!]
\includegraphics[width=6.3 cm]{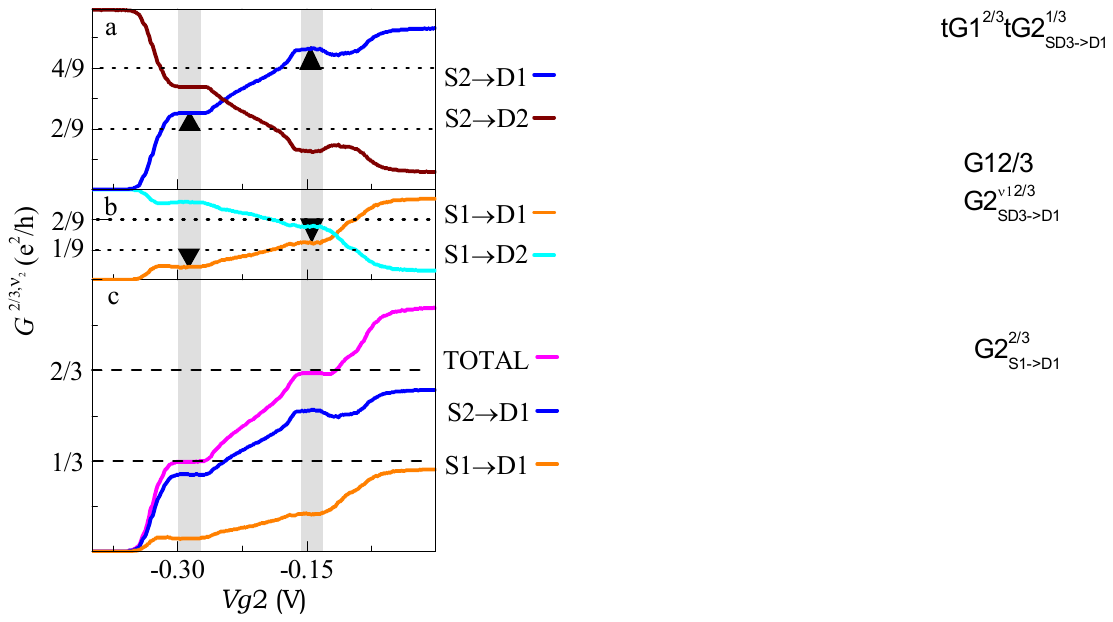}
\centering \caption[ ] 
{\label{fig:deviation} (color online) (a) Two-terminal combined conductance (TTCC) vs $Vg2$ at $\nu_1=2/3$ by exciting S2. Arrows indicate deviation of the TTCC (blue curve) measured at $\rm D1$ from the expected values (dotted lines) (b) TTCC vs $Vg2$ at $\nu_1=2/3$ by exciting S1. Arrows indicate deviation of the TTCC (orange curve) measured at $\rm D1$ from the expected values (dotted lines) (c) Replotting of blue and orange curves and plot of sum of the two (magenta). Gray strips indicate $\nu_2= 1/3$ (left) and 2/3 (right) region.}
\end{figure}

\begin{figure*}[htbp!]
\includegraphics[width=17 cm]{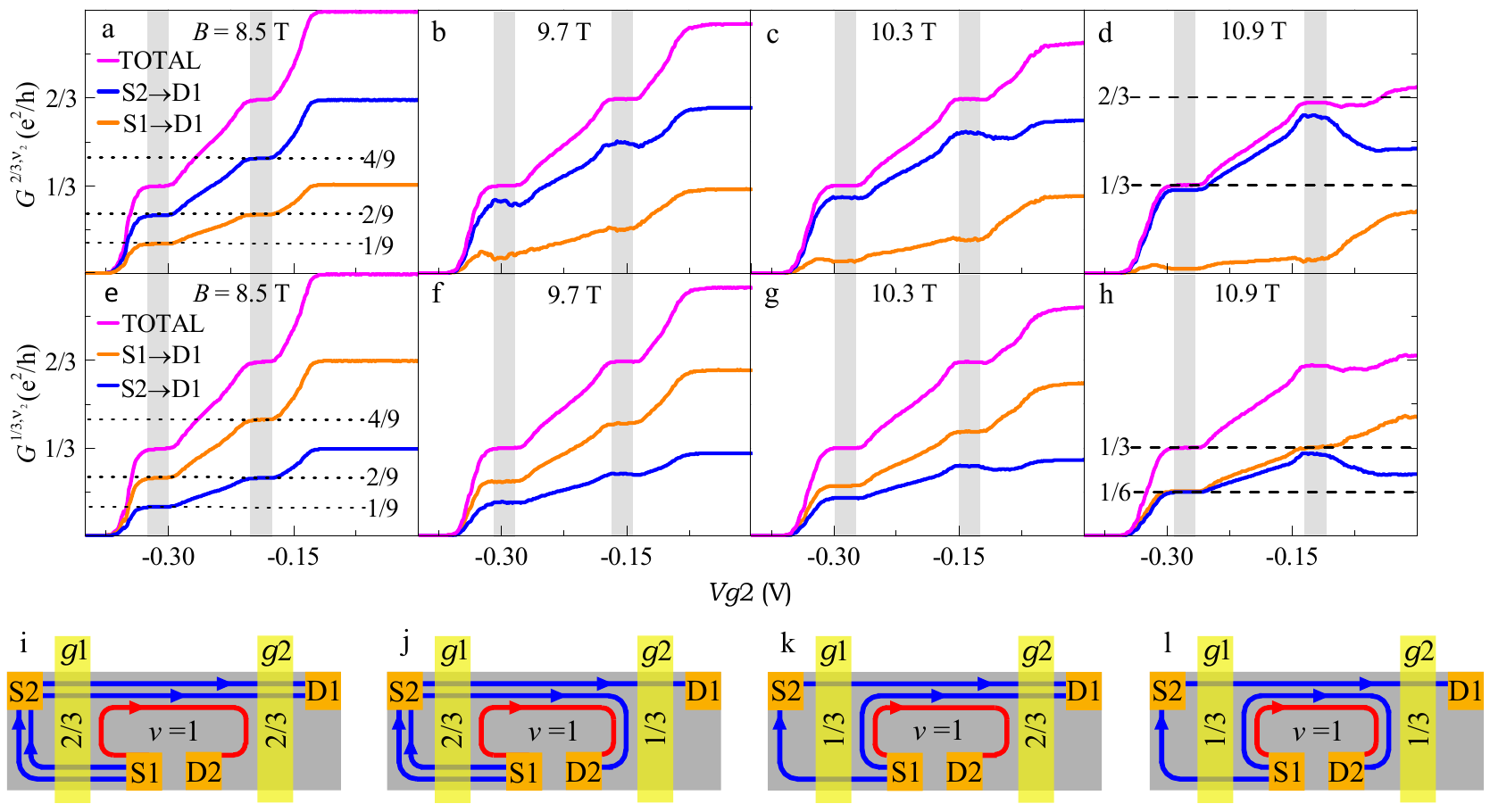}
\centering \caption[ ]
{\label{fig:limit} (color online) (a-d) Two-terminal combined conductance (TTCC) vs $Vg2$ at different $B$ fields when $\nu_1=2/3$. (e-h) TTCC vs $Vg2$ at different $B$ fields for $\nu_1=1/3$. Dotted lines in (a) and (e) show universal values of TTCC at 8.5 T. Dashed lines in (d) and (h) show new limits of TTCC at 10.9 T. Gray strips indicate $\nu_2= 1/3$ (left) and 2/3 (right) region. (i-l) Paths of three $1/3$ modes at different $\nu_1$ and $\nu_2$ values inferred from the measurements. Non current carrying modes are not shown.}
\end{figure*}

By setting $\nu=1$ and $\nu_1 = 2/3$, a $2/3  \ {\rm nA}$ current at $26 \  {\rm Hz}$ transmitted from S2 source and $1/3 \  {\rm nA}$ at $17 \ {\rm Hz}$ reflected current from source S1 are injected into the edge states along the top mesa boundary. Measured TTCCs are plotted as a function of $Vg2$ in Figs. \ref{fig:deviation}(a-b) at $B = 9.95 \ {\rm T}$. We see an enhancement of $G^{~2/3,\nu_2}_{\rm S2 \rightarrow D1}$ from expected universal conductance values defined by the filling fractions: $\nu_1\times\nu_2=4/9$ and $2/9$ (Fig. \ref{fig:deviation}(a), blue line) when $\nu_2$ is $2/3$ and $1/3$ respectively \cite{ButtikerPRL,ButtikerPRB,buttiker1992}. Also, we see suppression of $G^{~2/3,\nu_2}_{\rm S1 \rightarrow D1}$ from the expected universal values $(1-\nu_1)\times\nu_2=2/9$ and $1/9$ (Fig. \ref{fig:deviation}(b), orange line) when $\nu_2$ is $2/3$ and $1/3$ respectively. Conservation of current is evident from the compensating nature of the measured TTCCs in Figs. \ref{fig:deviation}(a and b). The TTCCs corresponding to the currents that reach the contact $\rm D1$ when injected from the sources $\rm S1$ and $\rm S2$ are plotted together in Fig. \ref{fig:deviation}(c). The sum of the two curves (magenta line) resembles the universal gate characteristics of $g2$ by compensating deviations of the TTCCs.

Deviation of the TTCCs from the expected universal limits depends on the magnetic field strengths. Figures \ref{fig:limit}(a-d) show the TTCCs similar to Fig. \ref{fig:deviation}(c) maintaining $\nu_1=2/3$ but at different magnetic fields within $\nu=1$ IQH plateau. At a field $B=8.5  \ {\rm T}$, the TTCCs reach the expected universal quantized conductance values (Fig. \ref{fig:limit}(a)). The TTCC values increasingly deviate from the universal limits (Figs. \ref{fig:limit}(b-d)) with increasing $B$ fields and tend to saturate at high field ($B = 10.9 \ {\rm T}$) to new TTCC limits tabulated in Table \ref{table1}. Similar deviation of TTCCs from the expected universal limits with increasing $B$ fields can be seen also maintaining $\nu_1=1/3$ (Figs. \ref{fig:limit}(e-h)) and the new TTCC limits are tabulated in Table \ref{table1}. These new TTCC limits at strong $B$ fields are the main observation of this letter.

These new TTCC limits cannot be explained by a simple picture with an integer edge mode. The observation of new TTCC limits suggests edge reconstruction at the natural mesa boundary consistent with Ref.~\onlinecite{Beenakker1990}, where three downstream $1/3$ charge modes arise from incompressible Laughlin like gaps \cite{Laughlin1983} corresponding to filling fraction 2/3 and 1/3 as the $\nu$ value reduces from the bulk value 1 to zero \cite{Chang1992}. 
Of these three, the two outer most $1/3$ charge modes are similar to the fractional edge modes of $2/3$ FQH state \cite{meir1994,sabo2017}, having charge equilibration length $l_r^o$ of the order of few micrometers. These two outer modes completely equilibrate with each other over propagation length $l = 125 \ {\rm \mu m}$ along the top boundary within $\nu = 1$ plateau. We assume that the innermost $1/3$ charge mode possess longer equilibration length $l_r^i \gg l$ at high $B$ field end and equilibrates with the others at low $B$ field end of $\nu=1$ plateau. With this model, the new TTCC limits at high $B$ and universal TTCC values at low $B$ can be explained using the schematic Figs. \ref{fig:limit}(i-l).

In Fig. \ref{fig:limit}(i) both the top-gated regions are at $2/3$ filling, where the two $1/3$ charge modes from $\rm S2$ reach $\rm D1$ without equilibrating with the innermost mode. As a result the TTCC reaches the new limit $G^{~2/3,2/3}_{\rm S2 \rightarrow D1}$  = 2/3 as observed in Fig. \ref{fig:limit}(d). In Fig. \ref{fig:limit}(j), out of two modes from S2, outermost $1/3$ charge mode reaches D1 and the other is reflected to D2, as a consequence the new TTCC limit $G^{~2/3,1/3}_{\rm S2 \rightarrow D1}$ = 1/3 is observed in Fig. \ref{fig:limit}(d). None of the modes from S1 reach D1 in the above two cases, resulting in the new TTCC limits of 
$G^{~2/3,2/3}_{\rm S1 \rightarrow D1}$ = $G^{~2/3,1/3}_{\rm S1 \rightarrow D1} \to 0$ as observed in Fig. \ref{fig:limit}(d). At lower magnetic field all three modes completely equilibrate leading to equipartition of the injected current and hence the TTCCs reach the universal values (Table \ref{table1}) as observed in Fig. \ref{fig:limit}(a).
\renewcommand{\arraystretch}{1.5}
\begin{table}
\centering
\begin{tabular}{p{3.5 cm}p{2.0 cm}p{2.0 cm}}
\hline
Two-terminal combined conductance (TTCC)  & Universal limit at low $B$  & New  limit at high $B$ \\
\hline
$G^{~2/3,2/3}_{\rm S2 \rightarrow D1}$  & 4/9  & 2/3 \\
$G^{~2/3,1/3}_{\rm S2 \rightarrow D1}$ & 2/9  & 1/3 \\
$G^{~2/3,2/3}_{\rm S1 \rightarrow D1}$ & 2/9  &  0 \\
$G^{~2/3,1/3}_{\rm S1 \rightarrow D1}$ & 1/9  &  0 \\
\hline
$G^{~1/3,2/3}_{\rm S2 \rightarrow D1}$ & 2/9  & 1/3 \\
$G^{~1/3,1/3}_{\rm S2 \rightarrow D1}$ & 1/9  & 1/6 \\
$G^{~1/3,2/3}_{\rm S1 \rightarrow D1}$ & 4/9  & 1/3 \\
$G^{~1/3,1/3}_{\rm S1 \rightarrow D1}$ & 2/9  & 1/6 \\
\hline 
\end{tabular}
\caption{Low and high field limits of conductances ($e^2/h$). \label{table1}}
\end{table}

When $\nu_1=1/3$ (Figs. \ref{fig:limit}(k-l)), the outermost $1/3$ charge mode from $\rm S2$ completely equilibrates with the middle $1/3$ mode along the top boundary. One (Fig. \ref{fig:limit}(l)) or both (Fig. \ref{fig:limit}(k)) of these two modes reach $\rm D1$ depending on $\nu_2$ values $1/3$ or $2/3$ respectively. As a consequence TTCCs reach to the new limits $G^{~1/3,1/3}_{\rm S2 \rightarrow D1}$ = 1/6 and $G^{~1/3,2/3}_{\rm S2 \rightarrow D1}$ = 1/3 (blue line, Fig. \ref{fig:limit}(h)). At $\nu_1=1/3$, the current in the middle $1/3$ charge mode injected from $\rm S1$ similarly equilibrates with the outermost mode. Again, one (Fig. \ref{fig:limit}(l)) or both (Fig. \ref{fig:limit}(k)) of these two modes reach to $\rm D1$ depending on $\nu_2$ values. Resulting new TTCC limits can be estimated to be $G^{~1/3,1/3}_{\rm S1 \rightarrow D1}$ = 1/6 and $G^{~1/3,2/3}_{\rm S1 \rightarrow D1} = 1/3$ (orange line, Fig. \ref{fig:limit}(h)) when $\nu_2$ is $1/3$ and $2/3$ respectively. At low $B$ fields (Fig. \ref{fig:limit}(e)) the TTCCs reach the universal conductance values (Table \ref{table1}). Using the model (Figs. \ref{fig:limit}(i-l)), all the TTCC can be represented in terms of the transmission probabilities \cite{ButtikerPRB} of the fractional edge modes \cite{SM2}.
\begin{figure}[h!]
\includegraphics[width=8.0cm]{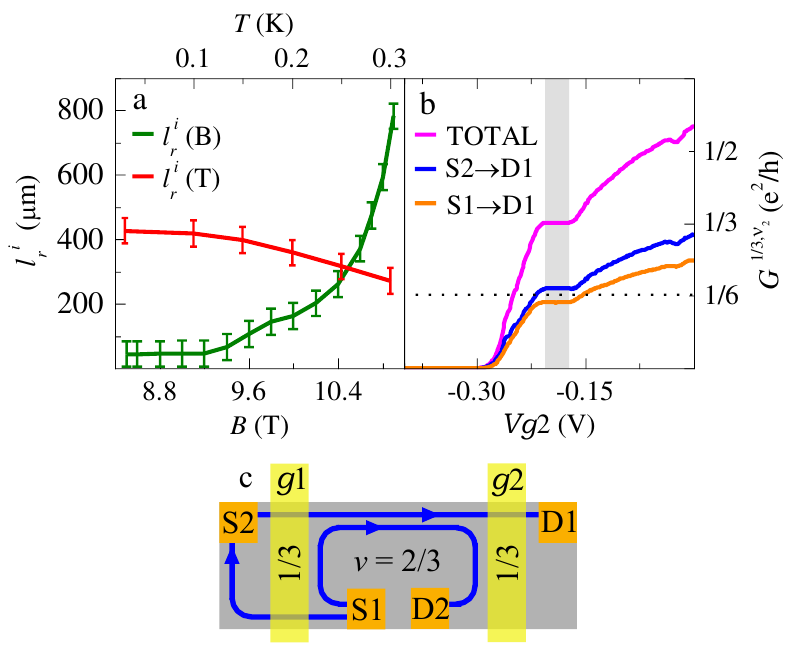}
\centering \caption[ ] 
{\label{fig:2by3} (color online) (a) Equilibration length of the innermost mode vs $B$  at $T = 30$ mK (green curve) and vs $T$ at $B=10.55$T (red curve). (b) Two-terminal combined conductance (TTCC) vs $Vg2$ at $\nu = 2/3$ for $\nu_1 = 1/3$. Gray strip indicates $\nu_2= 1/3$ region. Dotted line shows expected universal conductance limit for $\nu_1 =\nu_2 = 1/3$. (c) Schematic quantum Hall circuits for $\nu = 2/3$ and $\nu_1 = \nu_2 = 1/3$ based on two downstream 1/3 charge modes (blue lines).}
\end{figure}

Within the QH edge modes, energy is equilibrated by heat transfer \cite{altimiras2010,Lunde2010,rosenblatt2017} and charge is equilibrated by quasi-particle scattering \cite{Kouwenhoven1990,sabo2017}. In this experiment charge equilibration length $l_r^i$ of the innermost mode can be estimated from the current measured between S1 and D1 when $\nu_1=\nu_2=2/3$ (Fig. \ref{fig:limit}(i)). The current injected from S1 in the innermost mode is transferred by quasi-particle scattering into two outer modes over the propagation length $l$ \cite{SM2} that reach D1 and corresponding TTCC can be derive \cite{MullerEquilibration,Gafen2018,Lin2019,SM2} as:
\begin{equation}
G^{~2/3,2/3}_{\rm S1 \rightarrow D1} = \frac{2}{9} -  \frac{2}{9} e^{-3l/2l_r^i},
\label{equation:1}
\end{equation}
where the pre-factors are fixed by the boundary conditions - no scattering into outer modes at $l=0$ and  full equilibration at $l \gg l_r^i$. 
Figure \ref{fig:2by3}(a) (green line) shows the $B$ field dependence of the equilibration length estimated from variation of $G^{~2/3,2/3}_{\rm S1 \rightarrow D1}$ obtained from measurements similar to Figs. \ref{fig:limit}(a-d). Estimated equilibration length of the innermost mode increases from about zero to $777 \pm 40 \ \mu $m  with increasing $B$ field. Qualitatively $B$ field dependent equilibration length can be understood from decreased equilibration rate due to increase in separation between the modes resulting from shrinking of the $\nu=1$ region (Figs. \ref{fig:limit}(i-l)) with increasing $B$ field. For a fixed magnetic field $l_r^i$ generaly increases with lowering temperature and saturates to quantum scattering limit \cite{SM2} as shown in Fig. \ref{fig:2by3}(a) for $B=10.55$ T.

Increased equilibration length of the innermost mode with $B$ field (Fig. \ref{fig:2by3}(a)), suggests the possibility of higher equilibration length $l_r^o$ of the two outermost modes at high $B$ field beyond the $\nu = 1$ plateau. At higher magnetic field, inner most $1/3$ charge mode disappears as the bulk filling fraction reaches $\nu = 2/3$ and the outer most two 1/3 charge modes remain in the FQH system \cite{Beenakker1990,meir1994,sabo2017}. 
To examine the equilibration properties of the two $1/3$ charge modes at $\nu = 2/3$, we measure the TTCC from S1 to D1 by setting $\nu_1$ = $\nu_2 =1/3$ as shown in Fig. \ref{fig:2by3}(c). If fully equilibrated, half of the current form S1 should reach $\rm D1$ and the TTCC limit would be $1/6$. At the largest field $B = 13.98$ T accessible; we find TTCC value of $0.155 \pm 0.003$ instead of $1/6$ (Fig. \ref{fig:2by3}(b)) and this TTCC can be written as \cite{MullerEquilibration,Gafen2018,Lin2019,SM2}
\begin{equation}
G^{~1/3,1/3}_{\rm S1 \rightarrow D1} (\nu=2/3)=  \frac{1}{6} - \frac{1}{6}e^{-2l/l_r^o}.
\label{equation:2}
\end{equation} 
The equilibration length of the $1/3$ edge modes at $\nu = 2/3$ is increased to $l_r^o = 104 \pm 4 \ \mu $m at $B = 13.98$ T. 
The $l_r^o$ must be increasing within $\nu = 1$ plateau in micrometer range but much smaller $l_r^o$ results in full equilibration of the outer modes.

Increased equilibration lengths $l_r^i$ and $l_r^o$ of the FQH modes potentially results from high stability of incompressible FQH regions between the modes due to enhancement of Coulomb energy at higher $B$ fields. Quantitative analysis of magnetic field dependent equilibration of the 1/3 charge modes is left for future investigations.

Our model relies on adiabatic continuity of the edge modes of the FQH fluids beneath the gates $g1/g2$ into the FQH edge modes of the $\nu=1$ IQH fluid (Figs. \ref{fig:limit}(i-l)). Such adiabatic continuity is not expected if multiple modes coupled to one \citep{grivnin2014nonequilibrated} or a FQH edge mode partially reflected at this interface. Inhomogeneous carrier distribution \cite{karmakar2004effects} or high disorder \cite{Kouwenhoven1990} in samples can also prevent such adiabatic continuity. These complications are absent in our device, allowing quantized partition of $\nu = 1$ edge state.

In conclusion, we have demonstrated three fractional modes at a $\nu=1$ edge originating from dominated incompressibility of FQH states at filling 1/3 and 2/3 along the smooth mesa boundary. Our results suggest a possibility of finding more complex edge reconstruction in cleaner 2DEGs. Fractionalized IQH edge presents interesting new possibilities to explore fractional quasiparticle interferometry, tunneling, equilibration and statistics.

{\it Acknowledgements}: We thank K Sengupta, G. Baskaran and J K Jain for useful comments. We thank A Ratnakar for collaboration on a closely related calculation. GJS acknowledges support from DST-SERB grant ECR/2018/001781.

%\nocite{*}
\bibliography{Edgefractionalization}
\end{document}

\section*{{Supplemental Material: Magnetic field dependent equilibration of fractional quantum Hall edge modes}}
In the supplemental material, we discuss the experimental techniques, present further experimental data and analysis. In section \ref{sec:device}, we discuss the device structure and dimensions as well as the circuit diagram used in the experiment. In section \ref{sec:current}, we discuss the current distribution under the gate $g1$ to address formation of a uniform fractional fluid of filling $\nu_1$ = 1/3 or 2/3 underneath the gate. We discuss charge equilibration between the co-propagating fractional edge modes in section \ref{sec:equi} and the TTCC values are estimated as a function of equilibration length. In section \ref{sec:Tij}, TTCC is represented by transmission probabilities of the fractional edge modes and observed TTCC limits are explained in terms of transmission probabilities. In the last section emergent characteristics features of the charge equilibration length $l_r^i$ are discussed.

\section{Device structure, circuit diagram and details of measurements\label{sec:device}}

\begin{figure}[h!]
\centering
\includegraphics[width=15cm]{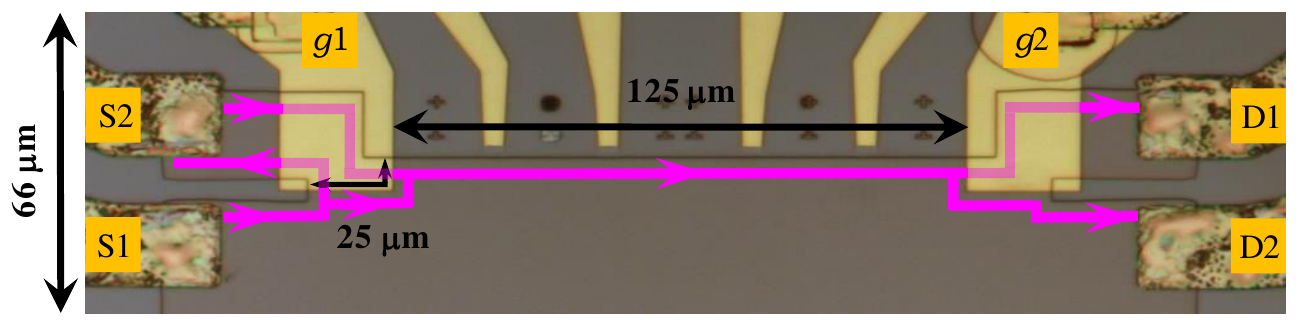}
 \caption{Optical image of the device structure. Pink lines denote current path in the mesa under the quantum Hall regime. Separation between the two gates $g1$ and $g2$ is $l = 125~\mu$m. Length of the gate boundary beside the central mesa region is $l_g = 25~\mu$m.}
\label{fig:device diagram1}
\end{figure}

\begin{figure}[h!]
\centering 
\includegraphics[width=15cm]{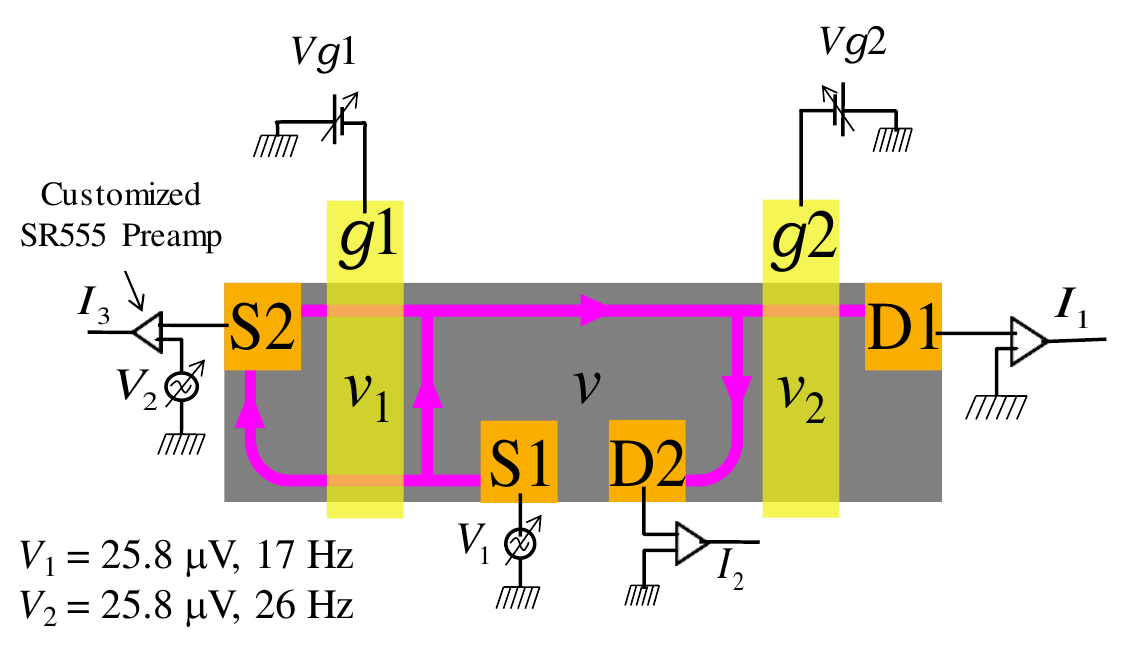}
\caption{Topologically equivalent device structure and measurement circuit diagram under quantized condition.}
\label{fig:device diagram2}  
\end{figure}

In this experiment, we use a one-sided modulation-doped AlGaAs/GaAs hetero-structure grown by molecular beam epitaxy. The two-dimensional electron gas (2DEG) resides at the AlGaAs/GaAs hetero-interface located $100$ nm below the top surface. A Si $\delta$-doped layer placed $42$ nm above the hetero-interface for modulation doping. The mesa is defined by wet chemical process and the Ohmic contacts are made by thermal deposition of metallic layers of Ni/AuGe/Ni/Au of thickness $10$/$100$/$10$/$100$ nm after photolithography. Ohmic contacts are annealed for a duration of $90$ s at 450$^o$C using a rapid thermal annealer. The gates are made by thermal evaporation of $10$ nm Ti followed by $100$ nm Au \cite{karmakar2011controlled}. 

The optical image of the device is presented in Fig. \ref{fig:device diagram1}. Size of Ohmic contacts (S1, S2, D1 and D2) on the mesa is approximately $20\times20~\mu$m$^2$. The contacts S1 and D2 are directly connected to the central mesa region. The device is symmetric about the bottom boundary of Fig. \ref{fig:device diagram1} and only the top half of the device is shown for simplicity. The topologically equivalent circuit diagram is shown in Fig. \ref{fig:device diagram2}.  The mesa width of the device is nearly 70 $\mu$m. The pink lines show the current paths in the device at filling fraction $\nu = 1$ in the quantum Hall (QH) regime. Ineffective current paths are not shown. From source S1, a fraction of current reaches contact S2 flowing underneath the gate $g1$ and rest of the current travel along the $g1$ gate boundary of length $l_g$ = 25 $\mu$m (Fig. \ref{fig:device diagram1}). Similarly a part of the current also travel along the $g2$ gate boundary of length $l_g$ = 25 $\mu$m to reach D2 contact and rest of the current reach D1 contact passing underneath the gate $g2$. The gate $g1$ controls the amount of current from S2 reaching the central mesa region passing through the top edge underneath the gate $g1$, as well as the amount of current reaching S2 from S1 passing along the bottom edge underneath the gate.

To set the central mesa region at filling $\nu=1$, we measure the two-terminal magneto-resistance (2TMR) between S1 and D2 \cite{karmakar2011controlled}. The output current at the contact D2 is measured with a voltage excitation at the S1 contact, keeping all other contacts S2 and D1 disconnected. Therefore, no current passes through the contacts S2 and D1. As a result, total current in the central meas region is unaffected by gate voltage $Vg1$ and $Vg2$. In particular, at filling $\nu=1$ of the central mesa region 2TMR must corresponds to $e^2/h$ conductance irrespective of gate voltage $Vg1$ and $Vg2$. During 2TMR measurement $Vg1$ and $Vg2$ is set to zero Volt. Inset-a of Fig. 1 of the main paper shows the 2TMR trace as a function of the applied magnetic field. This allows us to know the range of the magnetic fields at which a $\nu = 1$ plateau is formed in the central mesa region. We have not done any correction of contact/wire resistance in the 2TMR and conductance measurements because wire/contact resistance is much lower than the QH resistance 25.8 kOhm. Therefore in our sample, contact resistances are low enough that quantized conductance values are obtained without much error.

In Fig. \ref{fig:device diagram2}, the measurement circuit diagram is shown with a topologically equivalent device structure under quantized condition. At S2 contact a customized SR555 current-to-voltage pre-amplifier is connected to facilitate simultaneous measurement of output current from S1 to S2 and putting voltage excitation on the contact S2. Source contacts S1 and S2 are excited with $25.8~\mu$V at frequency 17 Hz and 26 Hz respectively. Output currents from D1 and D2 at frequency 17 and 26 Hz are measured by four lock-in amplifiers using suitable current to voltage pre-amplifiers. Output current at S2 having frequency 17 Hz is measured by a lock-in amplifier. During the experiment, we measure all the above five output currents simultaneously and always monitor the quantization condition of the fractional incompressible fluid beneath the gate $g1$ and IQH fluid in the central mesa region.

\section{Current distribution underneath the gate $g1$\label{sec:current}}

\begin{figure}[h!]
\includegraphics[width=12 cm]{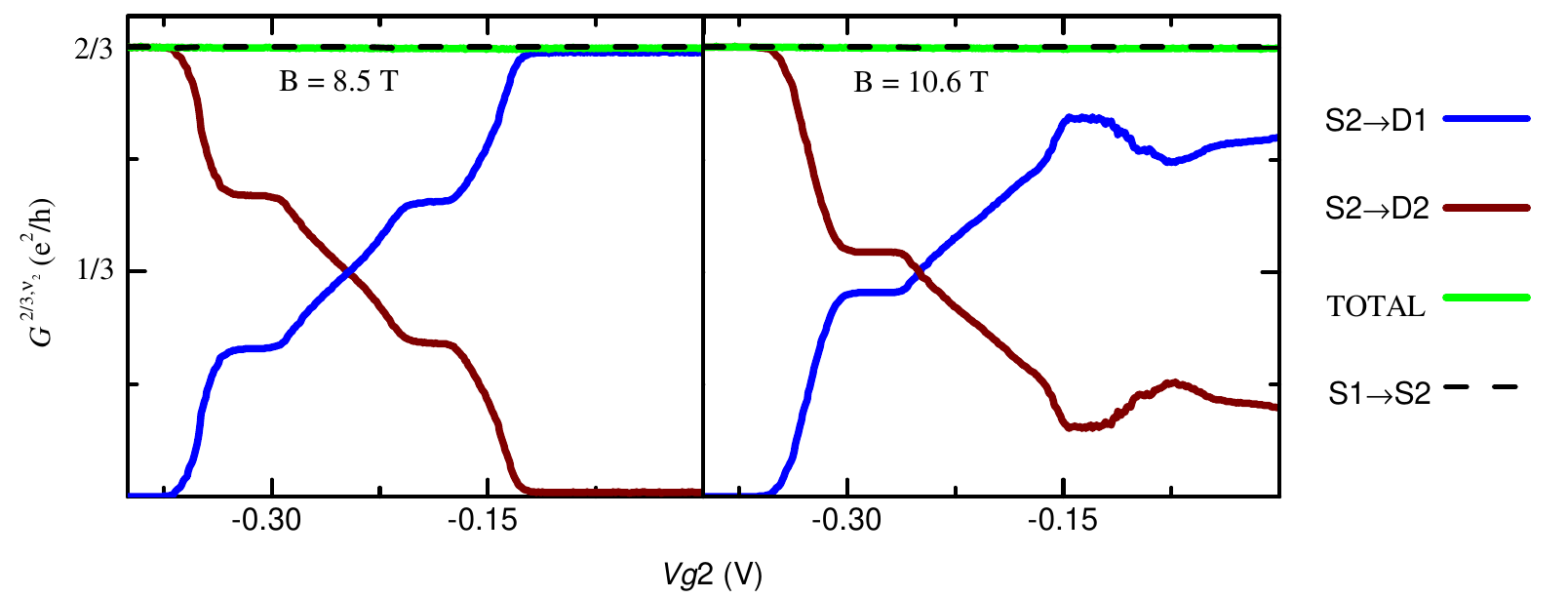}
\centering \caption{Two-terminal combined conductance (TTCC) vs $Vg2$ at two different magnetic fields when $\nu_1 = 2/3$. The TTCCs  measured at D1 and D2, in response to a 26 Hz voltage excitation at S2 are shown using the blue and wine curves respectively, similar to Fig. 2a of the main paper. The sum of the two TTCCs, indicated by the green curve is the total conductance of the top edge passing underneath the gate $g1$. Constancy of the top edge conductance (green line) indicates conservation of current with $Vg2$ sweep. Dashed line corresponds to the two-terminal transmitted conductance between S1 and S2, when S1 is excited by 17 Hz voltage signal.}
\label{fig:SMfig3} 
\end{figure}

\begin{figure}[h!]
\includegraphics[width=12 cm]{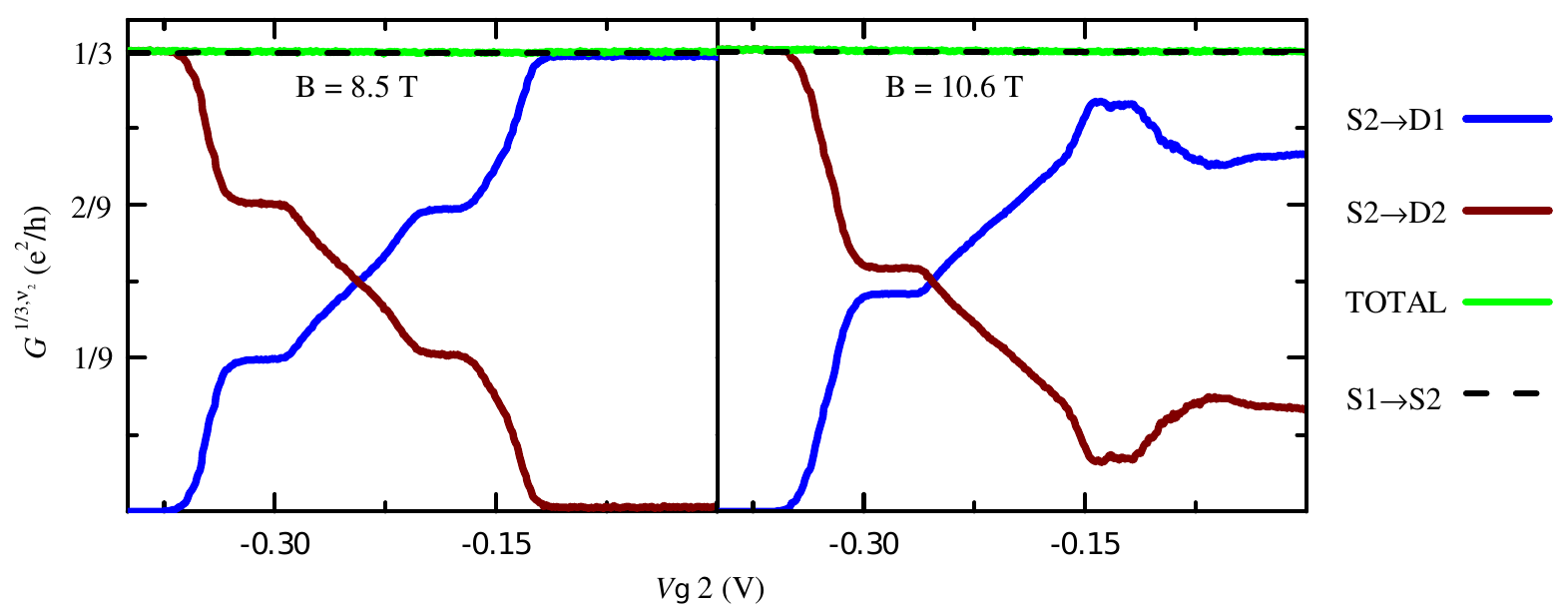}
\centering \caption{Two-terminal combined conductance (TTCC) vs $Vg2$ similar to Fig. \ref{fig:SMfig3}, at two different magnetic fields when $\nu_1 = 1/3$. The sum of the blue and wine curves is green line, which corresponds to the total conductance of the top edge underneath the gate $g1$. Dashed line corresponds to the conductance of the bottom edge underneath the gate $g1$.}
\label{fig:SMfig4} 
\end{figure}

At S2 contact a customized SR555 current to voltage pre-amplifier enables us to measure output current coming from S1 and passing through the bottom edge underneath the gate $g1$ when filling fraction $\nu_1$ is set at 1/3 or 2/3. Before taking data in Fig. 3 of the main paper, we always set the filling fraction  $\nu_1$ = 1/3 or 2/3 by measuring this conductance between S1 and S2, and accordingly $Vg1$ is adjusted (inset b of Fig. 1 of the main paper). During data acquisition of Figs. 3(a-h) of the main paper, we also monitor transmitted conductance S1 $\rightarrow$ S2, which is plotted (black dashed line) in Fig. \ref{fig:SMfig3} for the case of $\nu_1 = 2/3$ and in Fig. \ref{fig:SMfig4} for the case of $\nu_1=1/3$. The plots reveal that the conductance of the bottom and top edge modes underneath the gate $g1$ is same for any $Vg2$, any filling fraction $\nu_1$ = 1/3 or 2/3 and any magnetic field within the $\nu = 1$ QH plateau. Quantization of these conductances to 2/3 and 1/3 indicates a uniform FQH state formation under the gate $g1$, as any non-uniformity will lead to back scattering between the top and bottom edge modes beneath the gate, resulting in a deviation from the quantization.

\section{Equilibration along the edge\label{sec:equi}}

The experiment is conducted by selectively exciting a set of fractional modes and detecting same or another set of modes with electrostatic gating technique as described in the main paper. These populated modes co-propagate with other unpopulated modes along the top edge of the central region during which the different modes equilibrate transferring charges into the unpopulated modes. After equilibration between the modes the current in a subset of the modes are then measured at D1 and D2 contacts. The resulting two-terminal combined conductances (TTCCs) depend on the the extend of equilibration between the modes. Since equilibration happens along the top edge of the central mesa region, it forms the key physical length scale $l=125~\mu$m in the problem. In this section we describe the two-terminal combined conductance as a function of the length $l$.

We will first focus on equilibration processes between the two copropogating modes (labeled A and B) of identical conductance $\sigma$ via quasi-particle scattering/transfer. We follow and adapt the stochastic process of equilibration introduced in Ref.~\onlinecite{Muller1992}. We define a single parameter namely the equilibration length $l_r^o$ which corresponds to the average distance traveled by a quasi-particle between two inter-mode scattering event. The average number of inter-edge mode scattering events per unit length is $1/l_r^o$. Such an uncorrelated quasiparticle scattering model is suitable under small excitation voltage and weak disorder.

We consider the case where a current is injected only into mode $A$ at position $l=0$. The current in mode A at position $l$ along the channel is proportional to the probability $P(l)$ of finding each quasi-particle at a distance $l$ injected into mode A at $l=0$. Assuming incoherent scattering at the rate $1/l_r^o$ per unit length, the probability $P(l)$ evolve with $l$ as follows
\begin{equation}
P(l+dl)= P(l)\left [1-\frac{dl}{l_r^o} \right] + \left[1-P(l)\right]\frac{dl}{l_r^o},
\label{equation:1}
\end{equation}
where the two terms correspond to the charge retained in A and that scattered back from B to A over a short distance $dl$. The solution, with the initial condition $P(0)=1$, is
\begin{equation}
 P(l) = \frac{1}{2}(1+e^{-2l/l_r^o}).
\label{equation:2}
\end{equation}   
The current $\sigma V$ carried by the mode $A$ at $l=0$, is distributed between the modes A and B as $P(l)\sigma V $ and $(1-P(l))\sigma V $ respectively at position $l$. In our experiment at bulk filling fraction $\nu=2/3$, two edge modes have conductance $1/3$ each in unit of $e^2/h $. We excite one edge mode connected to contact S1 and measure the output current at D1 (see Fig. 4c of the main paper). TTCC between S1 and D1 is linked with the current transfer between the modes and hence the TTCC is given by (Eq. 2 of the main paper)
\begin{equation}
G^{~1/3,1/3}_{\rm S1 \rightarrow D1}(\nu=2/3)=\frac{1}{3} (1-P)=\frac{1}{6} - \frac{1}{6} e^{-2l/l_r^o}.
\label{equation:3}
\end{equation}

Based on the experimental results summarized in Figs. 3(a-h) and 4a of the main paper, we infer the existence of three fractional edge modes of conductance $\sigma = 1/3$ each around the bulk state of filling fraction $\nu=1$. The results also indicate that the two outer most modes have very short equilibration lengths and completely equilibrate with each other over the extent of $\nu=1$ IQH plateau. The inner mode posses finite equilibration length $l_r^i$ and equilibrates with the two outer modes. We extend the model described above to the present case of three channels and assume that equilibration of the inner mode with the outer ones happen through uncorrelated quasi-particle scattering. During the measurement of the conductance between S1 and D1 (orange line in Figs. 3(a-d) of the main paper), the current $\sigma V$ injected into the inner mode, partly scatters into the two outer modes at the top edge and is measured at D1. The probability $Q(l)$ that a quasi-particle injected at $l = 0$ in inner mode to be found at a distance $l$ in the same mode is governed by the condition
\begin{equation}
Q(l+dl)= Q(l)\left[1-\frac{dl}{l_r^i}\right] + \left[1-Q(l)\right]\frac{\alpha dl}{l_r^i}.
\label{equation:4}
\end{equation}
Here $1/l_r^i$ is the rate of scattering per unit length from the inner mode to the two outer modes and $\alpha/l_r^i$ is rate of scattering per unit length from one of the outer modes to the inner mode. Solving this with boundary condition $Q(0)=1$ and assuming steady state equilibrium at large $l$ to $Q(\infty)=1/3$ (current is divided equally in all modes), gives $\alpha = 1/2$ and 
\begin{equation}
Q(l)=\frac{1}{3}(1+2e^{-3l/2l_r^i}).
\label{equation:5}
\end{equation}
From above Eq. 5 the TTCC between S1 and D1 is given by (Eq. 1 of the main paper)
\begin{equation}
G^{~2/3,2/3}_{\rm S1 \rightarrow D1}=\frac{1}{3}(1-Q) = \frac{2}{9} - \frac{2}{9}e^{-3l/2l_r^i}.
\label{equation:6}
\end{equation}  
Moreover, the above stochastic model gives results identical to the model that is studied and verified in Ref.~\onlinecite{Lin2019} (also see Ref.~\onlinecite{Nosiglia2018}) with inter-channel scattering conductance per unit length. The exponential form describing the equilibration process has been extensively used for inter-mode charge transfer in IQH and FQH systems\cite{Muller1992,Wurtz2002,Ensslin2003,Takagaki1994,Paradiso2011} and the form have been verified with different sample lengths \cite{Muller1992,Takagaki1994,Paradiso2011}.

\section{Representation of TTCC by transmission probabilities\label{sec:Tij}}

In this section, we discuss two-terminal combined conductance (TTCC) in terms of transmission probabilities originating from equilibration between co-propagating fractional edge modes along the top boundary of the central mesa region. TTCC between two contacts can be expressed as a combination of transmission probabilities of fractional edge modes connecting the source and detector contacts. For the representation, the outer, middle and inner modes are labeled 1,2 and 3 respectively. Following B\"{u}ttiker approach \cite{Buttiker1988,Beenakker1990} in fractional edge modes, the TTCC can be expressed as 
\begin{equation}
G^{~\nu_1,\nu_2}_{\rm S \rightarrow D} = \frac{1}{3}\sum_{i;j} T_{ij},
\label{equation:7}
\end{equation} 
where $T_{ij}$ represents the probability for the transmission of a current from mode $i$ connected to the source into mode $j$ connected to the detector. The summation $i(j)$ is over all the modes in the top edge connected to the source S(D). Transmission probability $(T_{ij})$ between the respective modes depend on the charge equilibration over the co-propagation length  $l = 125~\mu$m and is symmetric $(T_{ij} = T_{ji})$. At high magnetic field end of the $\nu = 1$ plateau (Fig. 1 of the main paper), only outer two modes completely equilibrate, but inner most mode does not equilibrate with the outer modes, hence transmission probabilities are $T_{11} = T_{22} = T_{12} = T_{21} = 1/2$, $T_{33}=1$ and $T_{13} = T_{23}= T_{31} = T_{32} = 0$. At low magnetic field end of the plateau, all the three modes completely equilibrate with each other; hence all the transmission probabilities become 1/3. In the intermediate values of magnetic field, the transmission probabilities $T_{ij}$ varies from one limit to the other. 

In case of $\nu_1=\nu_2=2/3$ outer two modes connect S2 to D1 and inner mode connects S1 to D2 (Fig. 3i of the main paper). The corresponding TTCC can be expressed as:
\begin{align}
&G^{~2/3,2/3}_{\rm S2 \rightarrow D1} = \frac{1}{3}\sum_{i=1,2;j=1,2} T_{ij}=\frac{1}{3}(T_{11}+T_{12}+T_{21}+T_{22})\text{ and }\nonumber\\
&G^{~2/3,2/3}_{\rm S1 \rightarrow D1} = \frac{1}{3}\sum_{i=3;j=1,2} T_{ij}=\frac{1}{3}(T_{31}+T_{32}).
\label{equation:8}
\end{align}
Putting the values of $T_{ij}$, we get the TTCC values as tabulated in table 1 of the main paper. Similarly for all other combinations (Figs. 3(j-l) of the main paper), TTCC values can be estimated at the low and high magnetic field end of the $\nu = 1$ QH plateau as tabulated in table 1 of the main paper.

Now we discuss the resemblance of the gate characteristics to the total conductance shown in Figs. 3(a-h) of the main paper. In case of $\nu_1=\nu_2=2/3$, total conductance 2/3 of the top edge underneath the gate $g1$ (Fig. \ref{fig:SMfig3} here) can be expressed as
\begin{equation}
G^{~2/3,2/3}_{\rm S2 \rightarrow D1} + G^{~2/3,2/3}_{\rm S2 \rightarrow D2} = \frac{2}{3}.
\label{equation:9}
\end{equation}
From the symmetry of $T_{ij}$, we obtain
\begin{equation}
  G^{~2/3,2/3}_{\rm S2 \rightarrow D2} =\frac{1}{3}(T_{13}+T_{23}) = G^{~2/3,2/3}_{\rm S1 \rightarrow D1}.
\label{equation:10}
\end{equation}
The above two equations suggest following TTCC relation:
\begin{equation}
G^{~2/3,2/3}_{\rm S2 \rightarrow D1} + G^{~2/3,2/3}_{\rm S1 \rightarrow D1} = \frac{2}{3}.
\label{equation:11}
\end{equation}
The above relation (Eq. 11) indicates that the total conductance (measured by 26 Hz and 17 Hz excitation) should reach 2/3 for $\nu_1=\nu_2=2/3$ for any magnetic field within the $\nu=1$ QH plateau. This resembling of gate characteristic is solely arising from the symmetry of transmission probabilities $T_{ij}=T_{ji}$ and is experimentally verified in Figs. 3(a-d) of the main paper. This verification is the proof that the length $l$ is the key dimension of the experiment, since transmission probabilities $T_{ij}$ depends on the co-propagation length $l$ and equilibration lengths of the fractional edge modes. In this experiment, the quantity $ G^{~2/3,2/3}_{\rm S1 \rightarrow D1}=(T_{31}+T_{32})/3$ is measured by varying magnetic field (Figs. 3(a-d) of the main paper) and magnetic field dependent equilibration length $l_r^i$ of the inner most mode is estimated (Fig. 4a of the main paper).

\section{Characteristics of $l_r^i$\label{sec:lri}}

\begin{figure}[h!]
\includegraphics[width=12 cm]{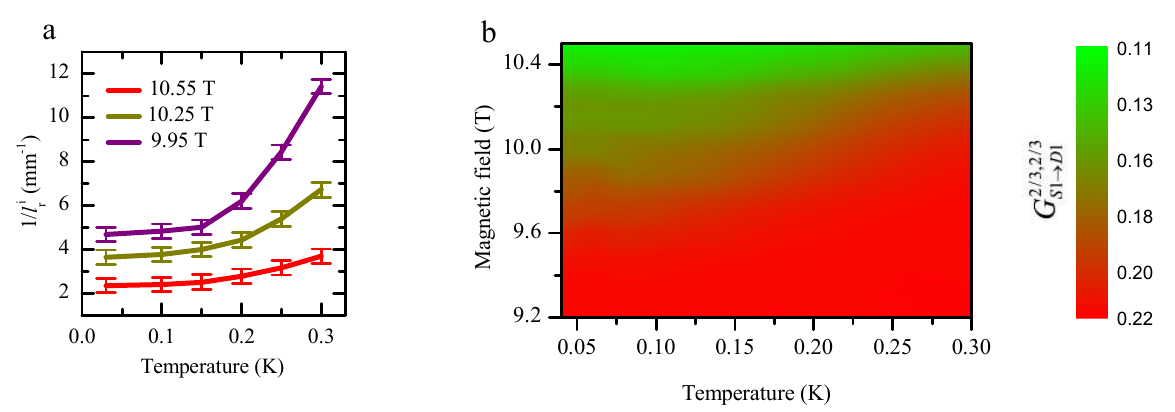}
\centering \caption{(a) Temperature dependence of $1/l_r^i$ at three different magnetic fields (b) Color plot of two terminal combined conductance (TTCC) between (S1 $\rightarrow$ D1) at $\nu_1=\nu_2=2/3$ as a function of magnetic field and temperature. Red shade indicates value closer to the universal conductance of $2/9$.}
\label{fig:SMfig5}   
\end{figure}     

Temperature dependence of charge equilibration, quantified by the inverse of the equilibration length $1/l_r^i$ of the innermost mode is plotted in Fig. \ref{fig:SMfig5}a as a function of $T$ for three different magnetic field values. The equilibration lengths are obtained by fitting the TTCC (between S1 and D1 when $\nu_1=\nu_2=2/3$) to Eq. 1 of the main paper. Equilibration rate decreases with lowering temperature and saturates at the lowest temperatures. The nature of the variation suggests a quantum scattering assisted charge  equilibration at low temperatures, and additional thermal scattering assisted equilibration as temperature increases. Increase of magnetic field reduces equilibration as seen in Fig. 4a of the main paper.

Color plot in Fig. \ref{fig:SMfig5}b shows the variation of the TTCC between S1 and D1 as a function of the magnetic field and temperature for $\nu_1=\nu_2=2/3$. Decreasing equilibration rate with increasing magnetic field and low temperatures is reflected in the deviation (green shade) of the conductance from the universal value $2/9$ (red shade).
 
%\nocite{*}
\bibliography{SMreference}